# Phen-DC$_3$ Induces Refolding of Human Telomeric DNA into a Chair-type Antiparallel G-quadruplex through Ligand Intercalation

Anirban Ghosh,[a] Marko Trajkovski,[b] Marie-Paule Teulade-Fichou,[c] Valérie Gabelica*[a] and Janez Plavec*[b,d,e]

[a]     Dr. A. Ghosh, Dr. Valérie Gabelica

CNRS, INSERM, ARNA, UMR 5320, U1212, IECB, Université de Bordeaux, F-33600 Pessac, France

E-mail : v.gabelica@iecb.u-bordeaux.fr

[b]     Dr. M. Trajkovski, Prof. Dr. J. Plavec

Slovenian NMR Centre, National Institute of Chemistry, Hajdrihova 19, 1000, Ljubljana, Slovenia

E-mail : janez.plavec@ki.si

[c]     Dr. M.-P. Teulade-Fichou

        Institut Curie, CNRS UMR 9187, Inserm U1196, Université Paris-Saclay, Orsay, France

[d]     Prof. Dr. J. Plavec

Faculty of Chemistry and Chemical Technology, University of Ljubljana, 1000, Ljubljana, Slovenia

[e]     Prof. Dr. J. Plavec

EN-FIST, Centre of Excellence, 1000, Ljubljana, Slovenia

        Supporting information for this article is given via a link at the end of the document.

**Abstract:** Human telomeric G-quadruplex DNA structures are attractive anticancer drug targets, but the target's polymorphism complicates the drug design: different ligands prefer different folds, and very few complexes have been solved at high resolution. Here we report that Phen-DC$_3$, one of the most prominent G-quadruplex ligands in terms of high binding affinity and selectivity, causes dTAGGG(TTAGGG)$_3$ to completely change its fold in KCl solution from a hybrid-1 to an antiparallel chair-type structure, wherein the ligand intercalates between a two-quartet unit and a pseudo-quartet, thereby ejecting one potassium ion. This unprecedented high-resolution NMR structure shows for the first time a true ligand intercalation into an intramolecular G-quadruplex.

**Introduction**

Alternate nucleic acid structures that differ from the typical Watson-Crick base-paired duplex have important biological roles.[1,2] In particular, G-quadruplexes (G4) are widespread in the human genome, as well as in plants, viruses, and bacteria.[3,4] A G4 structure is characterized by the formation of square planar guanine (G) quartets with Hoogsteen base-pairing.[5] Stacking between consecutive G-quartets is further stabilized by inter-quartet binding of monovalent or divalent cations.[6,7]

Recent bioinformatics and sequencing studies showed putative G4-forming sequences (> 700.000 PQS) within the human genome, while the estimated abundance of G4 structures in cells is lower but nonetheless impressive (around 10 000).[8] The distribution of G4s reveals regulatory roles in various cellular processes, including replication, recombination, transcription, translation, and telomere maintenance.[9–13] In parallel, deregulation related to G4 secondary structures adopted by G-rich sequences is linked to various human pathologies, especially cancers and neurological diseases, establishing G4s as crucial drug targets for novel therapeutic strategies.[14–16] In particular, stabilizing G4s with small-molecule ligands can inhibit telomerase or interrupt telomere capping and maintenance, resulting in cancer cell apoptosis.[17,18]

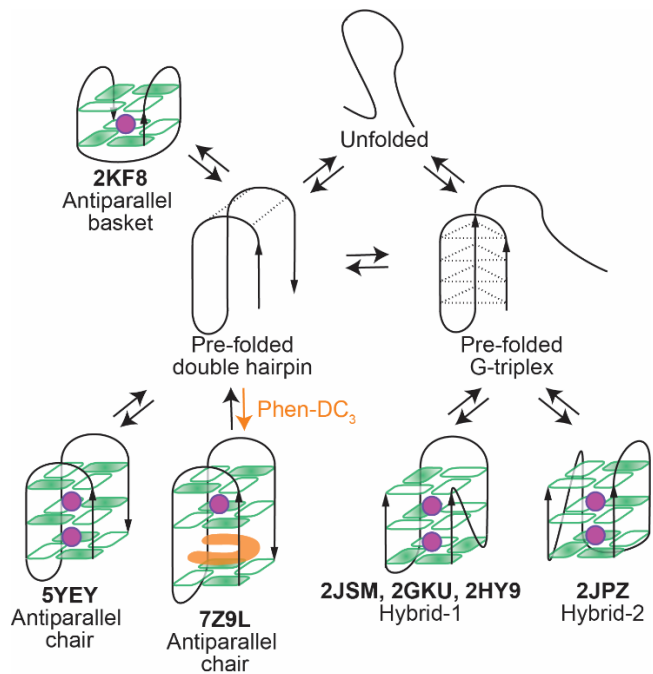

**Figure 1.** Polymorphism of the human telomeric G4s with variants in KCl solution and Phen-DC$_3$ induced stabilization of antiparallel chair-type structure. Schematic topologies of the three-quartet hybrid-1 (2JSM, 2GKU, 2HY9), the hybrid-2 (2JPZ) and the antiparallel chair (5YEY), the two-quartet antiparallel basket (2KF8)[19–23] and the complex between Phen-DC$_3$ and 23TAG in antiparallel chair conformation (this work, 7Z9L) are shown together with the double hairpin and G-triplex considered as intermediates and/or pre-folded conformations during G4 formation.[24] The annotations in bold correspond to the PDB entries of previously reported NMR structures. *Syn* and *anti*-guanines are denoted by green and white rectangles, respectively. The specifically bound K$^+$ ions in between consecutive G-quartets are denoted by magenta shaded circles.

However, G4 polymorphism makes it challenging to understand and design specific recognition of particular structure(s) by ligands.[25–27] The polymorphism arises from differences in strand stoichiometry, position and conformation of the loops connecting Gs and mutual orientations of G-tracts in the core of the structure.[28] The archetypal polymorphic sequence is the human telomeric repeat sequence (TTAGGG)$_n$, for which diverse (parallel, antiparallel and hybrid) intramolecular folding topologies have been identified (Figure 1).[22,24,29–34]

Structures of ligand complexes with hybrid-1, hybrid-2 or parallel forms of telomeric G4 have been documented by solution NMR (in K$^+$) and X-ray crystallographic studies,[35–40] but structural insights of ligand-induced stabilization of antiparallel G4 topology are still scarce in the literature. Yet this topology is one of the key G4 forms in a rugged folding free energy landscape,[23,41] and two-quartet basket-type G4s were reported both in cell-free and cellular systems under physiological conditions.[42,43] In all cases ligands were found to stack on terminal (5' and/or 3') G-quartet and stabilize the G4 in its initial folding topology. Typical examples are the hexaoxazole macrocycle L2H2-6M(2)OTD, or naphthalene diimides.[37,40] In a few cases (Au-oxo6, Epiberberine, NBTE), rearrangements of the 5'-end cap or loops surrounding the binding pocket have been reported.[35,36,38] However, ligand intercalation, although suggested by molecular modeling[44] and observed in-between GAGA and GCGC quartets of non-canonical VK2 structure,[45] has not been observed to date in G4 structures.

Phen-DC$_3$ is a U-shaped compound (Figure 2A), which binds G4s with nanomolar ($K_D$) affinity and can inhibit G4-unfolding by specialized helicases.[46–48] Phen-DC$_3$ is highly selective for G4s compared to duplex and single-stranded DNAs[49,50], and was previously shown to stack externally on the 5'-quartet of a parallel G4 adopted by the human c-myc promoter.[50] In contrast, V.G. showed that upon binding to four-repeat telomeric G4 sequences, Phen-DC$_3$ ejects one of the specific inner K$^+$ ion, suggesting that binding of Phen-DC$_3$ induced a structural transition from a three-quartet to a two-quartet topology.[49] Here we report the atomistic details of Phen-DC$_3$ binding to the hybrid-1 G4 adopted by the human telomeric sequence 23TAG, i.e. dTAGGG(TTAGGG)$_3$ (Figure 2B).[19,20] Our study reveals an unprecedented intercalative binding mode of Phen-DC$_3$ and a change of topology to antiparallel G4. The disclosure of the complex's high-resolution details at the DNA-ligand interface will expectedly be invaluable for optimizing G4-targeting strategies.

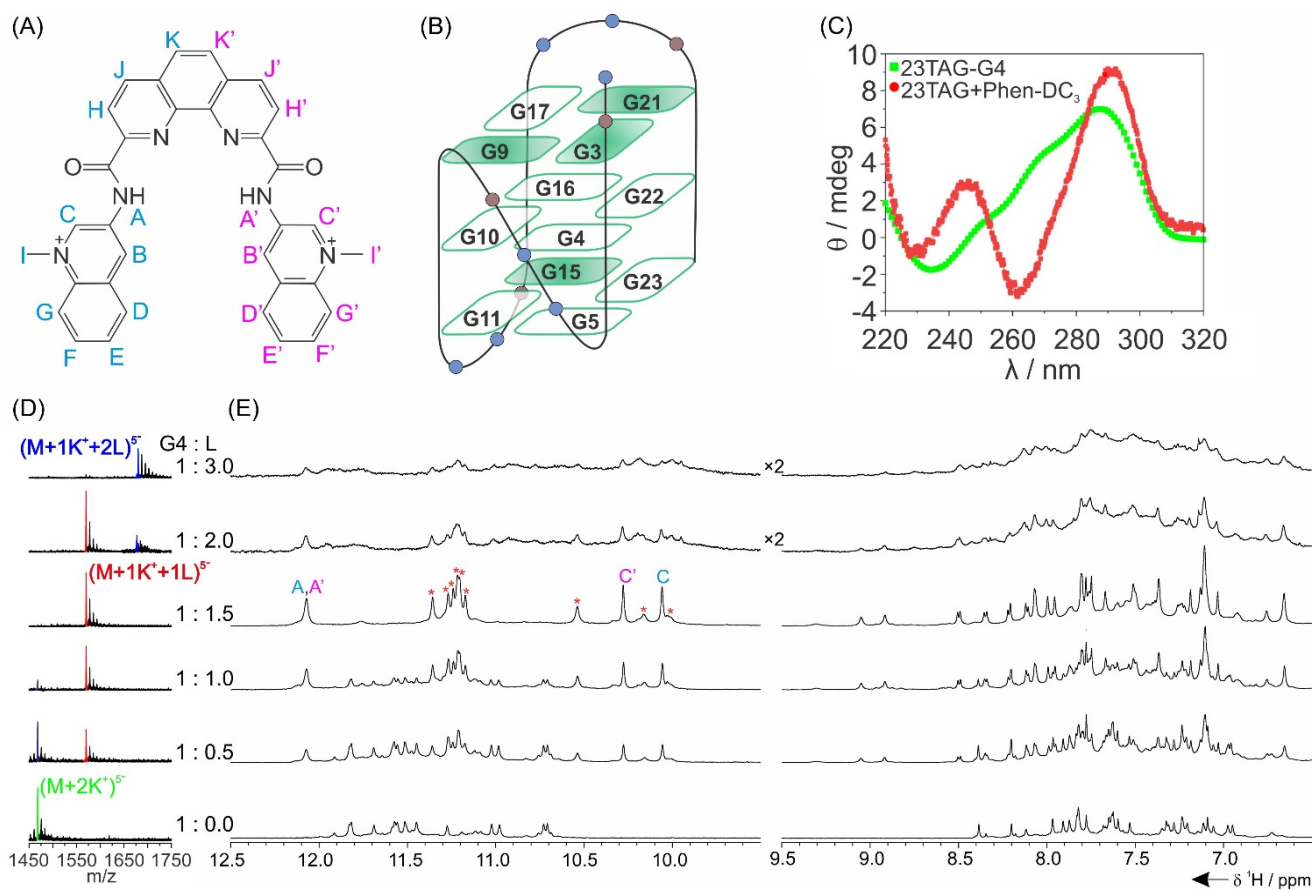

**Figure 2.** Interaction of 23TAG with Phen-DC$_3$. **(A)** Chemical structure of Phen-DC$_3$. Atom numbering and colors are depicted to indicate the asymmetry in the complex. **(B)** Hybrid-1 G4 adopted by 23TAG in the presence of K$^+$ ions. Green and white rectangles depict guanine residues exhibiting *syn* and *anti* glycosidic bond angle orientations, respectively. Blue and brown circles denote thymine and adenine residues, respectively. **(C)** CD spectra of 23TAG alone (green) and in the presence of 1.5 mole equivalent of Phen-DC$_3$ (red) at 25 °C in 70 mM KCl and 20 mM potassium-phosphate buffer (pH 7.0), at 10 µM oligonucleotide concentration. **(D-E)** ESI-MS and imino and aromatic regions of $^1$H NMR spectra of 23TAG titrated with Phen-DC$_3$, with the DNA:ligand ratio shown between the two panels. M stands for the oligonucleotide, L for the ligand, K$^+$ for potassium ion. In **(D)** M+2K$^+$ (green), M+1K$^+$+1L (red) and M+1K$^+$+2L (blue) indicate the major ESI-MS peaks for 0, 1 and 2 ligands bound, respectively. The ESI-MS spectra were recorded in 100 mM TMAA buffer and 1 mM KCl at 10 µM oligonucleotide concentration. In **(E)** the $^1$H NMR signals corresponding to the amide group and the nearby quinolinium moieties (left) in the 23TAG+PhenDC$_3$ complex are indicated with capital letters, while the guanine imino protons are indicated with red asterisks. The vertical intensities of the imino spectral region are twice those of the aromatic region. The vertical intensities of the NMR spectra at ratios of 1:2.0 and 1:3.0 are increased by a factor of 2 compared to the lower ratios shown below. $^1$H NMR spectra were recorded in 95% H$_2$O/5% $^2$H$_2$O, at 25 °C, 0.5 mM oligonucleotide strand concentration, 70 mM KCl and 20 mM potassium-phosphate (pH 7.0).

**Results and Discussion**

CD spectral changes of 23TAG upon binding to Phen-DC$_3$ in 100 mM KCl solution reveal a structural transition from a hybrid to an antiparallel topology (Figure 2C). ESI-MS titration in 1 mM KCl and 100 mM trimethylammonium acetate shows that the 23TAG+Phen-DC$_3$ complex has only one intra-G4 bound K$^+$ ion (Figure 2D). Up to 1.5 equivalent ligand concentration, the 1:1 (23TAG:Phen-DC$_3$) stoichiometry predominates. A 1:2 complex appears at higher molar ratios, still with one specifically bound K$^+$ ion. The same CD behavior was obtained in the ESI-MS buffer.

Without the ligand, the $^1$H NMR spectrum of 23TAG at physiological K$^+$ ion concentration exhibits two sets of twelve imino signals (Figures 2E, S1 and S2). The well-resolved imino signals of the major form correspond to the hybrid-1 structure reported previously,[19,20] coexisting with a minor hybrid-2 form.[19,51–53] Both hybrid structures have three G-quartets. Upon addition of 0.5 equivalents of Phen-DC$_3$, an additional set of imino $^1$H NMR signals appears, indicating slow inter-conversion on the NMR chemical shift timescale between the free G4s and a single 23TAG+Phen-DC$_3$ complex structure (Figure 2E). Upon increasing the ligand-to-G4 ratio to 1.5, the signals corresponding to the complex intensify and become sharper, while the signals of the free G4 structures gradually disappear. The complex exhibits eight sharp imino signals. We observe the same NMR signals in the ESI-MS buffer (Figure S2). Thus, the MS and NMR data support a new, striking binding mode of only two stacked G-quartets with one K$^+$ ion in-between. $^1$H NMR and CD also suggest that a similar complex is formed in the presence of Na$^+$ ions (Figure S3), as well as in KCl concentrations below 1 mM (see supplementary ESI-MS, CD and $^1$H NMR results, Figures S4-S5). Further addition of ligand up to 3 equivalents leads to broadening of the (imino) $^1$H NMR signals, indicating the formation of complex(es) with higher ligand-to-G4 stoichiometries.

In-depth structural characterization of the 1:1 23TAG+Phen-DC$_3$ complex was performed at 1.5 molar equivalents of ligand, for optimal spectral quality. The imino (H1), aromatic (H2, H6 and H8), sugar and methyl (thymine and N-methyl) NMR resonances of the 23TAG+Phen-DC$_3$ complex were assigned using $^{15}$N- and $^{13}$C-edited HSQC spectra on partially residue-specifically labeled 23TAG (Figures 3A-B, Figures S6-S8 and Table S1). Unambiguous assignment was complemented with 2D $^1$H-$^1$H NOESY, TOCSY and natural abundance $^1$H-$^{13}$C HSQC spectra (Figures 3C-D, Figures S9-S10 and Table S2-S3). Results show that the eight sharp imino signals in 1D $^1$H NMR spectra of the complex correspond to G4, G10, G16 and G22, and G5, G9, G17 and G21 (Figure 3A). These residues are involved in H-bonding within the two stacked G-quartets. Interestingly, two additional – albeit broader – imino resonances for G15 and G23 at ~10.28 and 10.06 ppm suggest the formation of a third, more dynamic G-quartet in the 23TAG+Phen-DC$_3$ complex, wherein the inherent protons of G3 and G11 would be much more exposed to solvent exchange.

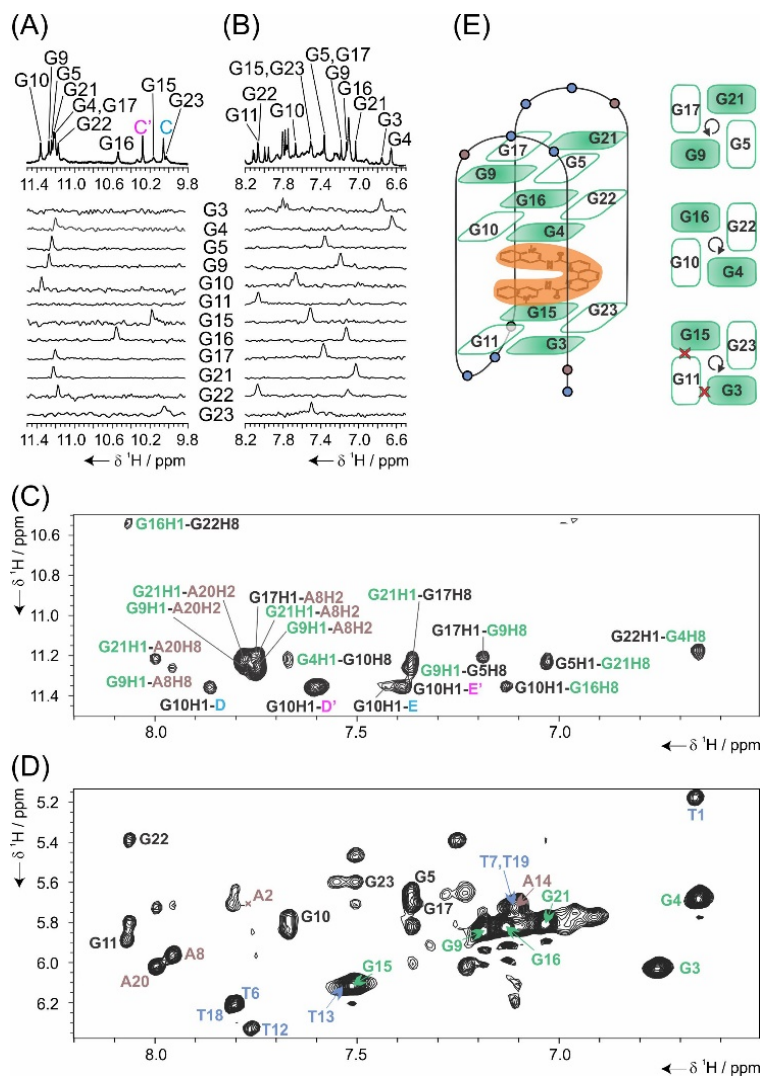

**Figure 3.** NMR analysis of G4 folding topology in the 1:1 23TAG+Phen-DC$_3$ complex. Unequivocal assignment of guanine **(A)** imino and **(B)** aromatic $^1$H NMR resonances of the complex by using 1D $^{15}$N- and $^{13}$C-edited HSQC spectra. The corresponding regions of $^1$H NMR spectrum are shown at the top together with indicated assignment of guanine as well as C and C' quinolinium protons signals. Plots of 2D $^1$H-$^1$H NOESY spectrum recorded at 300 ms mixing time showing **(C)** intra-quartet guanine H1-H8 correlations consistent with G4→G10→G16→G22 and G5←G9←G17←G21 H-bonding directionalities. Guanines exhibiting *syn* and *anti* glycosidic bond angle conformations are marked in green and black, respectively. Additionally, cross-peaks between guanine imino and adenine aromatic protons and between G10 H1 and Phen-DC$_3$ protons are indicated. **(D)** Medium-to-weak and strong intra-residual H1'-H8 cross-peaks for guanines exhibiting *anti* (black) and *syn* (green) conformations, respectively. Intra-residual H1'-aromatic NOEs of thymine (blue) and adenine residues (brown) are also shown. **(E)** Antiparallel chair-type G4 topology of the 23TAG+Phen-DC$_3$ complex, with H-bonding directionalities within G-quartets shown on the right. Two red x's indicate the absence of the respective imino proton signals in the $^1$H NMR spectra for G3 and G11. The ligand is highlighted in orange. NMR spectra were recorded in 95% H$_2$O/5% $^2$H$_2$O, at 25 °C, 70 mM KCl, 20 mM K$^+$-phosphate buffer (pH 7.0) and (A-B) 0.2 mM oligonucleotide concentration per strand and 0.3 mM ligand; (C-D) 0.5 mM oligonucleotide concentration per strand and 0.75 mM ligand.

The observed imino-aromatic correlations in NOESY spectra of the 23TAG+Phen-DC$_3$ complex are consistent with the following H-bond directionalities within the consecutive G-quartets: G4→G10→G16→G22 and G5←G9←G17←G21 (Figure 3E). Additionally, the G15H1-G23H8 and G23H1-G3H8 NOE interactions are very weak, suggesting formation of the third G3→G11→G15→G23 quartet (Figure S11). The strong and weak intra-nucleotide H1'(i)-H8(i) cross-peaks in NOESY spectra denote that G3, G4, G9, G15, G16 and G21 are in *syn*, while others predominantly adopt an *anti* conformation (Figures 3D and S10). Using the data described above, we could outline an antiparallel chair topology of G4 in the 23TAG+Phen-DC$_3$ complex (Figure 3E). Thorough analysis of cross-peaks in the NOESY spectra resolved several inter-quartet interactions, which define relative orientations of the constitutive residues, structural parameters of G4 (*e.g.*, roll, tilt, helical twist, and rise) and stacking between G4-G10-G16-G22 and G5-G9-G17-G21 quartets. Hydrogen-deuterium exchange NMR experiments showed longer protection (>3 days) of the imino protons of G4-G10-G16-G22 quartet from solvent exchange compared to the outer G-quartets (<1 hr), suggesting a central position in the complex (Figure S12).

The lack of sequential G-to-G NOE correlations suggests the absence of direct G4-G10-G16-G22 and G3-G11-G15-G23 stacking. This corresponds to the Phen-DC$_3$ intercalation site (Figure 3E). We call here the G3-G11-G15-G23 quartet a "pseudo-quartet" due to the absence of H1-H8 NOE correlations for G3-G11 and G11-G15 pairs. Notably, the calculated molecular model (*vide infra*) are consistent with all the guanines arranged in G-quartets, amongst which the G3-G11-G15-G23 quartet is more dynamic.

Several cross-peaks in the NOESY spectra, in particular the interactions of G5 with A8 and G17 with A20, demonstrate the close packing of some residues of the first and third loops (T7-A8 and T19-A20) with the adjacent G5-G9-G17-G21 quartet (Figures 3C and S10). NOEs were observed between G5 and T6, and G17 and T18, but these show that the two thymines are located in the grooves and not at the top of the G-quartet. The highly symmetric character of the two lateral loops connecting G5 to G9 and G17 to G21 in the 23TAG+Phen-DC$_3$ complex is suggested by near isochronous $^1$H NMR chemical shifts of aromatic and methyl protons of T6 and T18 as well as of T7 and T19.

Phen-DC$_3$ in its free form is symmetric and exhibits isochronous $^1$H NMR signals for A/A', B/B', etc. pairs of protons (Figure S13). Upon binding, however, distinct $^1$H NMR resonances are observed in the complex for all stereochemically related ligand protons (Figure 2 and Figure S14), except for the amide protons (A and A') that exhibit isochronous signal at ⌐~12.07 ppm, possibly due to exchangeable nature of NH protons. To assign ligand resonances, we analyzed a series of 2D NMR experiments (Figures 3-4 and Figure S15). These results, together with the four inter-quinolinium NOE correlations (B-B', B-D', D-D' and B'-D), are consistent with Phen-DC$_3$ positioned in an asymmetric environment, in a conformation that facilitates stacking with the neighboring G-quartets. The characteristics of the ligand confined in a non-uniform environment are further illustrated by the $^1$H NMR chemical shift changes, Δδ of 0.2 ppm for the N-methyl groups (I and I'), which could be unambiguously assigned by $^{13}$C-edited HSQC spectra of the 23TAG+Phen-DC$_3$ complex using the $^{13}$C-(N-methyl) labeled ligand prepared as previously described (Figure S8).[50] The particular orientation and conformation of Phen-DC$_3$ in the complex is consistent with the absence of NOE interactions between the two N-methyl groups.

The intermolecular NOE correlations reveal an interesting topology with a quartet-quartet stacking of opposite-polarity (G5-G9-G17-G21 and G4-G10-G16-G22) followed by a ligand-quartet block in which Phen-DC$_3$ is sandwiched between successive guanines characterized by same-polarity stacking, although the G-tracts are in antiparallel orientation (Figure 4 and Figure S15). Phen-DC$_3$ is intercalated, with its quinolinium units positioned between G15 and G16 on one side, and G10 and G11 on the other. The phenanthroline ring is located between G4-G22 and G3-G23 hydrogen bonded base pairs in the 23TAG+Phen-DC$_3$ complex.

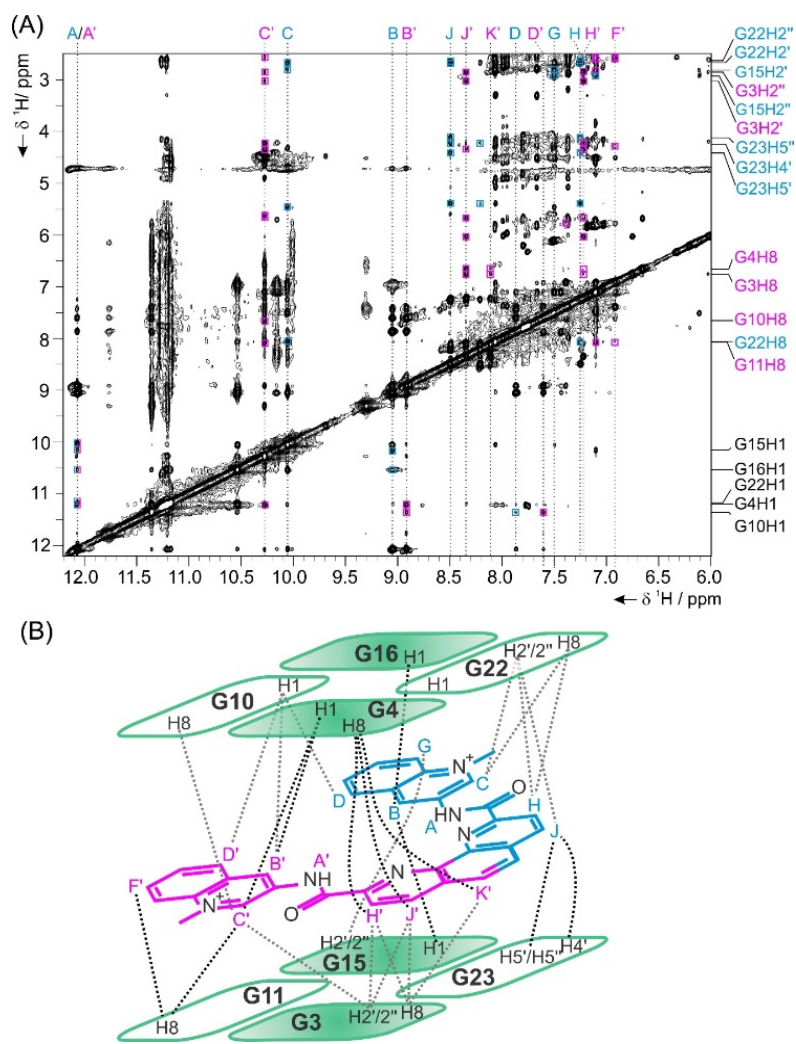

**Figure 4.** Inter-molecular interactions within the 23TAG+Phen-DC$_3$ complex. **(A)** The region of 2D $^1$H-$^1$H NOESY spectrum showing inter-molecular cross-peaks, amongst which key interactions are schematically depicted in **(B)**. A more completely annotated NOESY spectrum is presented in Figure S15. The spectrum was recorded at mixing time of 300 ms, in 95% H$_2$O/5% $^2$H$_2$O, at 25 °C, 0.5 mM oligonucleotide strand concentration, 0.75 mM ligand, 70 mM KCl and 20 mM K$^+$-phosphate buffer (pH 7.0).

Using the intra- (417 DNA-DNA and 4 ligand-ligand) and inter-molecular (73 DNA-ligand) NOE-derived distance restraints (Table S4), we calculated the solution structure of the 1:1 23TAG+Phen-DC$_3$ complex (Figure 5) by simulated annealing. The ten lowest energy structures of the 23TAG+Phen-DC$_3$ complex show an all-atoms R.M.S.D. of 0.675 Å (Table S5). The 10 structures from 50-ns unrestrained MD give an R.M.S.D. of 1.316 Å (Figure S16). The calculated high-resolution structure of the 23TAG+Phen-DC$_3$ complex shows four strands connected by three lateral TTA loops in an antiparallel chair-type G4 topology (Figure 5). G4-G10-G16-G22 and G5-G9-G17-G21 quartets exhibit an opposite-polarity 5-ring stacking, consistent with the CD profile, with inter-quartet distance of ca. 3.8 Å, and *nwnw* (*n*=narrow, *w*= wide) groove width combination (Figure S17). The two lateral T6-T7-A8 and T18-T19-A20 loops bridge the two narrow grooves and exhibit extensive stacking of their residues on the G5-G9-G17-G21 quartet (Figure 5). A8 and A20 are almost coplanar and form a continuous stack from the G5-G9-G17-G21 quartet to the T7-T19 base pair. The close positions of T7 H3 and T19 O4 atoms, as well as the T19 H3 and T7 O4 atoms, suggest the formation of a T-T base pair, with their imino protons in fast exchange with the solvent as indicated by the lack of the corresponding $^1$H NMR signals. On the other hand, T6 and T18 protrude into the wide grooves of the G4 with their respective arrangement, indicating potential hydrogen-bonding between their carbonyl oxygen atoms and the amino groups of G5 and G17.

The most interesting details of the high-resolution structure relate to Phen-DC$_3$, intercalated between the G4-G10-G16-G22 and G3-G11-G15-G23 quartets. The two quinolinium units are located between the nucleobases of the G10*anti*-G11*anti* and G15*syn*-G16*syn* steps and show a pronounced stacking with the purine moieties. Interestingly, the conformation of the intercalated Phen-DC$_3$ is non-planar with an angle of about 10° between the quinolinium ring planes, presumably to optimize stacking interactions with the neighboring G-quartets, which is supported also by unrestrained MD calculations. Both quinolinium rings are slightly tilted out of plane with respect to the phenanthroline ring, with the larger angle observed for the one between G15 and G16. N-methyl groups I and I' both point towards the narrow grooves, but their orientations with respect to the nearby strands are different. Moreover, the N-methyl group I is closer to the phosphodiester group of G15-G16 than G22-G23 step (5.2 Å vs. 9.0 Å, respectively), whereas on the other side the distances between I' and the phosphodiester groups of G3-G4 and G10-G11 steps are more similar (*i.e.,* 6.1 and 6.8 Å, respectively).

Compared to the previously published structure with Phen-DC$_3$ externally stacked on the parallel c-myc G4,[50] the position of the phenanthroline ring does not appear optimal with respect to stacking with the adjacent G3-G23 and G4-G22 base pairs. It is pushed towards the groove, slightly closer to the G22-G23 than to the G3-G4 step (Figure 5). The average distances between Phen-DC$_3$ and G4-G10-G16-G22 and G3-G11-G15-G23 quartets are 3.0 and 3.1 Å, respectively. These distances are shorter than the typical quartet-quartet distances, and indicate a close quartet-ligand-quartet packing, in line with the quartet formation being promoted by Phen-DC$_3$.[54] Notably, the G3-G11-G15-G23 quartet is capped by a T13-A2-A14 base triad with A2 serving as H-bond donor and acceptor to both T13 and A14 (Figure 5). T1 is stacked over A2 and caps the T13-A2-A14 base triad, which is consistent with the long-range NOE interactions observed between protons of T1 and T13, and T1 and A14. On the other hand, T12 is oriented towards the wide groove formed by the G9-G11 and G15-G17 strands. Notably, the T13-A2-A14 base triad is positioned at the edge of Hoogsteen-sides of G3, G15 and G23, leaving G11 more exposed to bulk water molecules. This local structural feature, together with the closer position of phenanthroline unit above G23 rather than G3, can explain why [1]H NMR signals are observed for imino protons of G15 and G23, but not for G11 and G3.

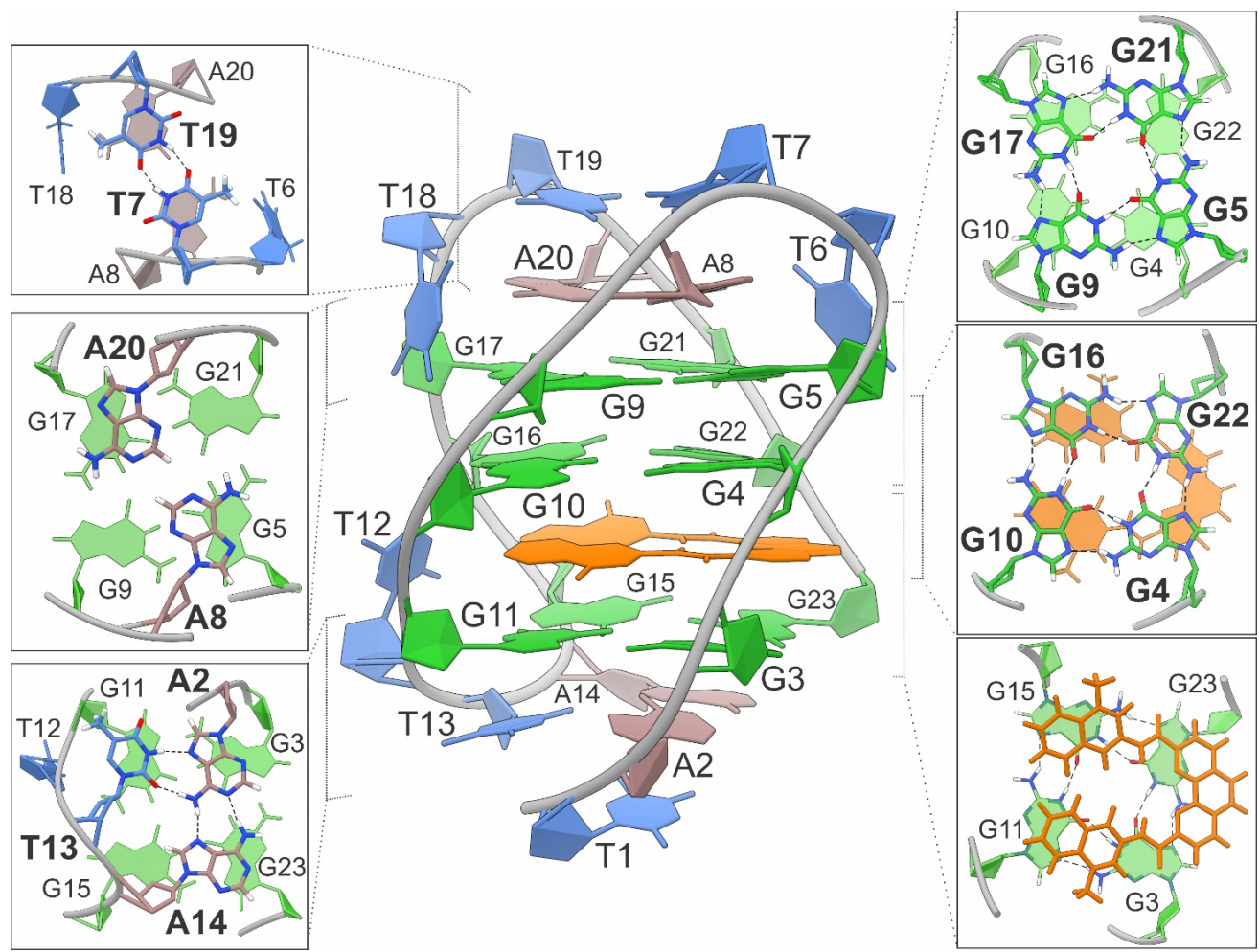

**Figure 5.** Solution-state structure of 1:1 23TAG+Phen-DC$_3$ complex. The lowest energy, high-resolution structure of the complex in the center is supplemented with structural details of the T6-T7-A8 and T18-T19-A20 loops (upper left). Top view show T7 and T19 in base-pair stacked above the consecutive adenine residues. Top view of A8 and A20 stacked on the nearby G5-G9-G17-G21 quartet (center left). Stacking of T13-A2-A14 base triad on G3-G11-G15-G23 quartet (bottom left). Top views on consecutive stacks of G5-G21-G17-G9 and G4-G10-G16-G22 (upper right), G4-G10-G16-G22 and Phen-DC$_3$ (center right), and Phen-DC$_3$ and G3-G11-G15-G23 quartet (lower right). The dashed lines are used to depict H-bonds.

Additionally, higher flexibility of this pseudo-quartet is supported by unrestrained 50 ns MD simulation results, in particular by the RMSD values of 0.945, 0.995 and 1.394 Å for G5-G9-G17-G21, G4-G10-G16-G22 and G3-G11-G15-G23 quartets, respectively. The capping of G3-G11-G15-G23 quartet by the T13-A14 and 5'-end overhanging residues (T1, A2) together with the structural juxtaposition of the first and the second loops noted above, indicates that the lateral TTA loops play a critical role in stabilizing the antiparallel chair-type G4 topology in the complex.

In the 23TAG+Phen-DC$_3$ complex, the two quinolinium rings are tilted with respect to the phenanthroline moiety, extending their stacking to the nearby Gs, consistent with the importance of flexible linkers connecting aromatic groups in a ligand.[37,50] The binding interface details furthermore suggest that the formation of 23TAG+Phen-DC$_3$ complex might be guided by quinolinium rings, rather than phenanthroline unit interactions. This notion is supported by the six-membered ring size and the strong electron-acceptor character of quinolinium units both being critical for maximizing π-π overlap with G-quartets and favoring stacking with guanine rings as with Phen-DC$_3$ analogs.[50,55,56] In fact, the position of the phenanthroline unit at the edge of the groove formed by the first and the last G-tract, i.e. G3-G5 and G21-G23, observed herein is reminiscent of the detail in the c-myc G4-Phen-DC$_3$ complex, where Phen-DC$_3$ is end-stacked on the 5' G-quartet with all guanines in *anti* conformation.[50] This earlier study proposed to design more effective Phen-DC$_3$ analogues by introducing modifications to the N-methyl groups (I and I'), J, J', K, and K'. Our structural findings suggest that to target the telomeric structure specifically, the design should focus on the latter two.[57]

In recent years, NMR studies on human telomere repeats demonstrated that a two-quartet basket-type G4 topology exists both in cell-free and cellular systems under physiological conditions.[42,43] This topology is one of the crucial G4 forms in the rugged folding energy landscape.[23,41] Recently, Frelih *et al.* identified the pre-folded G-triplex (pH 5.0) and double-hairpin (pH 7.0) adopted by 23TAG, providing their structural characteristics by NMR, which are in line with previous MD and time-resolved NMR studies.[24,53,58] Interestingly, the 23TAG+Phen-DC$_3$ complex scrutinized herein and the double-hairpin are of similar shapes with predisposition for H-bonding observed in the former. Considering its large aromatic surface of the three rings and U-shape, Phen-DC$_3$ has the capacity to interact with four Gs and thus may induce their assembly into G-quartets *via* a templating effect. This feature of Phen-DC$_3$ together with its bis-cationic charge is likely to promote ligand interactions in the unfolded/pre-folded states, even with minimal amounts of K$^+$ ions. We tested this hypothesis and indeed found that Phen-DC$_3$ binds to the pre-folded hairpin species and folding intermediates, as revealed by the evident changes in the $^1$H NMR imino region upon adding the ligand to a solution of 23TAG in the absence of intentionally added cations (Figure S18). This indicates that the complex formation does not necessarily require a fully pre-folded G4 structure, and supports the mechanism presented in Figure 1.

In line with this mechanism, a recent smFRET study reported no direct Phen-DC$_3$ induced conversion from hybrid to antiparallel telomeric G4 but rather ligand trapping of dynamically populated short-lived conformations.[27] Another study based on MD simulations suggested that conformational interconversion involves full unfolding of the hybrid structure and groove width reconfiguration.[59] Our results suggest that Phen-DC$_3$ traps 23TAG in a transient form and prevents further structural changes (e.g., refolding to major hybrid-1 structure), thus driving the equilibrium towards the complex observed herein. All the above evidence supports a conformational selection mechanism over the induced fit for Phen-DC$_3$ binding to 23TAG.

**Conclusion**

In summary, the high-resolution NMR structure presented here shows the structural details of a 1:1 complex wherein the ligand Phen-DC$_3$ is intercalated into an antiparallel chair-type G4 adopted by the telomeric sequence 23TAG. Most remarkably, the ligand templates the assembly of the third G-quartet through π-π stacking, and this third G-quartet is further stabilized on the other side by capping of the central lateral loop and the 5'-overhang. This creates an unprecedented intercalation site in-between two G-quartets. We suggest that transient pre-folded species adopted by human telomeric sequences may be the substrates for binding Phen-DC$_3$. The redistribution of folding topologies induced by Phen-DC$_3$ also underlines the limitation of traditional rationalization of ligand binding to G4,[60] based on stacking with pre-existing fully folded G4 structures.

Further work will be necessary to understand the driving force for the preferential binding of Phen-DC$_3$ to this antiparallel chair-type topology, which had never been reported for telomeric G4. In living cells, G4 polymorphism is expectedly highly complex, featuring fluctuations between several folded and unfolded substrates, yet to be explored in detail.[61–66] The disclosed chaperonic activity of Phen-DC$_3$ is invaluable as it highlights the ligand-induced formation of specific complex from unfolded or pre-folded G-rich sequences.[67] It thus provides a starting point for further engineering Phen-DC$_3$ analogs targeting not just all G4s, but specific G4 structures *in vitro* and *in vivo*.

**DATA AVAILABILITY**

The atomic coordinates and the list of chemical shifts of the 23TAG+Phen-DC$_3$ complex have been deposited in the Protein Data Bank and Biological Magnetic Resonance Data Bank with the accession code 7Z9L and 34714, respectively.

**Acknowledgements**

This work was financially supported by the European Union (H2020-MSCA-IF-2017-799695-CROWDASSAY), and benefited from access to NMR, MS and CD at the Plateforme de BioPhysico-Chimie Structurale of the IECB. V.G. and A.G. thank Dr. Frédéric Rosu for assistance and useful discussions. M.T. and J.P. acknowledge financial support from the Slovenian Research Agency [grants P1-0242 and J1-1704]. M.P.T.F. acknowledges financial support from the Fondation pour la Recherche Médicale (DCM20181039571). The authors acknowledge the CERIC-ERIC consortium for the access to experimental facilities and financial support as well as Dr. Laurent Sevaille (Institut Curie) for the synthesis of $^{13}$C-labelled Phen-DC$_3$.

**Keywords:** G-quadruplex • ligand intercalation • Phen-DC$_3$ • NMR spectroscopy • mass spectrometry